\documentclass[aps,english,prd,twocolumn,superscriptaddress,nofootinbib,preprintnumbers,10pt,notitlepage,eqsecnum]{revtex4}
\usepackage[latin1]{inputenc}
\usepackage{graphicx}
\usepackage{amsfonts}
\usepackage{amsmath,amsthm,amssymb}
\usepackage{enumitem}
\usepackage{bm}
\usepackage{graphicx}
\usepackage{amsfonts}
\usepackage{dcolumn}
\usepackage{color}
\usepackage{mathrsfs}

\def\be{\begin{equation}}
\def\ee{\end{equation}}
\def\ba{\begin{eqnarray}}
\def\ea{\end{eqnarray}}
\def\bs{\begin{subequations}}
\def\es{\end{subequations}}

\newcommand{\s}{{\cal S}}
\newcommand{\Z}{{\cal Z}}
\newcommand{\U}{{\cal U}}

\usepackage{color}

\makeatother

\usepackage{babel}
\makeatother

\begin{document}

\title{Inflationary gravitational waves in the effective field theory of modified gravity}

\author{Antonio De Felice}
\affiliation{Yukawa Institute for Theoretical Physics, Kyoto University, 606-8502, Kyoto, Japan}

\author{Shinji Tsujikawa}
\affiliation{Department of Physics, Faculty of Science, Tokyo University of Science, 1-3,
Kagurazaka, Shinjuku, Tokyo 162-8601, Japan}

\date{\today}

\begin{abstract}

In the approach of the effective field theory of modified gravity, we derive 
the second-order action and the equation of motion for tensor perturbations 
on the flat isotropic cosmological background.
This analysis accommodates a wide range of gravitational theories
including Horndeski theories, its generalization, and the theories 
with spatial derivatives higher than second order (e.g., Ho\v{r}ava-Lifshitz gravity). 
We obtain the inflationary power spectrum of tensor modes 
by taking into account corrections induced by higher-order spatial 
derivatives and slow-roll corrections to the de Sitter background. 
We also show that the leading-order spectrum in concrete modified 
gravitational theories can be mapped on to that in General Relativity 
under a disformal transformation.  
Our general formula will be useful to constrain inflationary 
models from the future precise measurement of the B-mode 
polarization in the cosmic microwave background.

\end{abstract}

\maketitle




\section{Introduction}
\label{intro} 

The detection of primordial gravitational waves in the Cosmic Microwave 
Background (CMB) offers an exciting possibility for probing the physics around 
the GUT scale. In particular, the inflationary 
paradigm \cite{Sta80,Kazanas,newinf} predicts the generation of 
nearly scale-invariant primordial tensor and scalar perturbations 
with the tensor-to-scalar ratio $r$ less than the order 
of 0.1 \cite{GWper,oldper}. 
Since the spectral index $n_s$ of scalar perturbations was measured 
by the Planck satellite in 
high precision ($n_s=0.9603\pm 0.0073$ at 68 \%\, CL \cite{Planck}), 
the precise bounds on $r$ from the ongoing and upcoming 
CMB B-mode polarization experiments \cite{QUIET,BICEP,Polar,Lite} 
are the next important step for approaching the origin of inflation. 

Many of the single-field inflationary models proposed so far belong to 
a class of Horndeski theories \cite{Horndeski}-- the most general 
Lorentz-invariant scalar-tensor theories with second-order equations of motion. 
In fact, the leading-order power spectra of tensor and scalar perturbations 
were derived for inflationary models in the framework of 
Horndeski theories \cite{KYY,GaoSteer}. 
These results were employed to place observational constraints on 
individual models (such as slow-roll inflation \cite{newinf}, 
k-inflation \cite{kinf}, Starobinsky inflation \cite{Sta80}, 
Higgs inflation \cite{Higgs,Germani}) from the 
WMAP and Planck data \cite{Reza,Kuro}.

There exist more general modified gravitational theories beyond 
the Horndeski domain.
Choosing the so-called unitary gauge in which the perturbation of 
a scalar field $\phi$ on the flat Friedmann-Lema\^{i}tre-Robertson-Walker 
(FLRW) background vanishes, the Horndeski Lagrangian can be expressed 
in terms of geometric scalar variables appearing in the 3+1 
Arnowitt-Deser-Misner (ADM) decomposition of space-time \cite{Piazza}. 

In Horndeski theories the coefficients in front of such ADM scalars have 
two particular relations, but it is possible to consider extended theories with 
arbitrary coefficients: Gleyzes-Langlois-Piazza-Vernizzi (GLPV) 
theories \cite{Gleyzes}. 
In general space-time the equations of motion in GLPV theories should 
contain derivatives higher than second order, but the Hamiltonian analysis 
based on linear perturbations on the flat FLRW background 
shows that GLPV theories have one scalar 
propagating degree of freedom without ghost-like Ostrogradski 
instabilities \cite{Gleyzes,Lin,GlHa,GaoHa}.
This second-order property also holds for the odd-type perturbations 
on the spherically symmetric background \cite{KaGe}. 

The full action of GLPV theories cannot be 
generally mapped to that of Horndeski theories \cite{GlHa,Tsu14} 
under the so-called disformal transformation \cite{Beken,Garcia}, 
so the two theories are not equivalent to each other. 
It is possible, however, to derive the two non-Horndeski pieces in the GLPV 
Lagrangian separately from a subset of the Horndeski Lagrangian 
under the disformal transformation.

Another example outside the Horndeski domain is 
Ho\v{r}ava-Lifshitz gravity \cite{Horava}, in which an anisotropic scaling 
in time and space plays a role for the realization of 
a power-counting renormalizable theory.  
In this case there are spatial derivatives up to 6-th order, with which 
Lorentz invariance is explicitly broken. 
The building blocks of Ho\v{r}ava-Lifshitz gravity are the three-dimensional 
ADM geometric scalars invariant under a foliation-preserving diffeomorphism. 

The effective field theory (EFT) of cosmological perturbations 
is a powerful framework to deal with low-energy degrees of freedom 
in a systematic and unified way \cite{Cremi,Cheung,Weinberg}. 
This approach is not only useful to parametrize
higher-order correlation functions of curvature perturbations generated during 
inflation \cite{Smith,Cremi2,Planck} but also to perform a systematic study 
for the physics of a late-time 
cosmic acceleration induced by the modification of gravity \cite{Quin}-\cite{Kase14}. 
In fact, recent studies \cite{Bloom,Piazza,Gleyzes,Gergely,Gao,RT14} showed 
that the EFT approach can encompass a wide range of theories including 
Horndeski/GLPV theories and Ho\v{r}ava-Lifshitz gravity.
 
The EFT approach of Ref.~\cite{Piazza} is based upon
a general Lagrangian $L$ in unitary gauge that depends on the 
lapse $N$ and several ADM geometric scalars constructed from 
the extrinsic curvature $K_{\mu \nu}$ and the three-dimensional 
intrinsic curvature ${\cal R}_{\mu \nu}$. The action expanded up to 
second order in scalar metric 
perturbations shows that the theory has one scalar degree of freedom 
with spatial derivatives higher than second order in general, while 
the time derivatives are of second order.
Both Horndeski and GLPV theories satisfy conditions for the absence 
of such higher-order spatial derivatives \cite{Piazza,Gergely,Tsuji14}.

The original projectable version of Ho\v{r}ava-Lifshitz gravity, in which 
the lapse $N$ depends on the time $t$ alone, is plagued by 
the strong coupling problem \cite{ins1,ins2}.
In the non-projectable version where $N$ depends on both time 
and space, the acceleration vector $a_i=\nabla_i \ln N$ does not vanish 
and hence several scalar quantities like $\eta\,a_i a^i$ can be present
in the Lagrangian ($\eta$ is a constant) \cite{Sergey}. 
In this case there are some parameter spaces of $\eta$ in which 
the strong coupling problem in the original theory can be alleviated. 
This strong coupling still remains an open issue 
without realizing a truly renormalizable and UV complete 
theory \cite{Soti,Sergey2}.  

The non-projectable version of Ho\v{r}ava-Lifshitz gravity
can be incorporated in the EFT approach of Ref.~\cite{Piazza} 
by taking into account additional geometric 
scalar quantities (associated with spatial derivatives up to 6-th order) 
to the Lagrangian \cite{Gao,RT14}. 
In Ref.~\cite{Kase14} the second-order action for scalar perturbations 
was derived for the generic EFT Lagrangian 
encompassing Horndeski/GLPV theories and Ho\v{r}ava-Lifshitz gravity.
This result can be useful for the computation of the primordial 
scalar power spectrum generated during inflation and for discussing 
conditions under which the ghosts and instabilities are absent 
(see Ref.~\cite{Tsuji14} for a review).

In this paper we employ such a general EFT approach 
for the study of tensor perturbations on the flat FLRW 
background. Our analysis is more generic 
than those of Refs.~\cite{Tsuji14,Gao,Cre} in that higher-order 
spatial derivatives appearing in Ho\v{r}ava-Lifshitz gravity 
are explicitly taken into account for the computation of 
the second-order action of tensor perturbations.
Unlike Ref.~\cite{Wands}, we do not consider the terms 
associated with the broken spatial diffeomorphism.
We provide a general formula for the inflationary power spectrum 
of tensor modes by taking into account slow-roll 
corrections to the leading-order spectrum.

This paper is organized as follows.
In Sec.~\ref{theorysec} we present the action of our EFT approach 
and briefly review how several modified gravitational theories 
are incorporated in our general framework.
In Sec.~\ref{persec} we derive the second-order action and 
the equation of motion for tensor perturbations.
In Sec.~\ref{gwspesec} we obtain the spectrum of gravitational 
waves generated during inflation and in Sec.~\ref{appsec} 
we apply the results to concrete modified gravitational theories.
Sec.~\ref{conclude} is devoted to conclusions.


\section{The general EFT action of modified gravity}
\label{theorysec} 

The EFT of cosmological perturbations is based upon 
the 3+1 decomposition of space-time described by 
the line element \cite{ADM}
\be
ds^{2}=g_{\mu \nu }dx^{\mu }dx^{\nu}
=-N^{2}dt^{2}+h_{ij}(dx^{i}+N^{i}dt)(dx^{j}+N^{j}dt)\,,  
\label{ADMmetric}
\ee
where $N$ is the lapse function, $N^i$ is the shift vector, and
$h_{ij}$ is the three-dimensional spatial metric. 
Introducing a unit vector $n_{\mu}=(-N,0,0,0)$ orthogonal to the constant 
$t$ hypersurfaces $\Sigma_t$, the induced metric $h_{\mu \nu}$ 
on $\Sigma_t$ can be expressed of the form $h_{\mu \nu}=g_{\mu \nu}
+n_{\mu}n_{\nu}$.

The extrinsic curvature is defined by 
$K_{\mu \nu}=h^{\lambda}_{\mu} n_{\nu;\lambda}
=n_{\nu;\mu}+n_{\mu} a_{\nu}$, 
where a semicolon represents a covariant derivative and  
$a_{\nu} \equiv n^{\lambda}n_{\nu;\lambda}$ 
is the acceleration vector. 
The scalar quantities that can be constructed from the 
extrinsic curvature are the trace of $K_{\mu \nu}$ and 
the square of $K_{\mu \nu}$, i.e.,  
\be
K \equiv {K^{\mu}}_{\mu}\,,\qquad
\s \equiv K_{\mu \nu} K^{\mu \nu}\,.
\ee

The internal geometry of $\Sigma_t$ is characterized 
by the three-dimensional Ricci tensor 
${\cal R}_{\mu \nu}={}^{(3)}R_{\mu \nu}$, 
which is dubbed the intrinsic curvature. 
{}From ${\cal R}_{\mu \nu}$ we can construct
the following scalar quantities:
\be
{\cal R} \equiv
{{\cal R}^{\mu}}_{\mu}\,,\qquad
\Z \equiv {\cal R}_{\mu \nu}
\mathcal{R}^{\mu \nu}\,, \qquad
\U \equiv \mathcal{R}_{\mu \nu}K^{\mu \nu}\,. 
\ee
Since it is possible to express the Riemann tensor 
$R_{\mu \nu \lambda \sigma}$ in terms of the Ricci tensor 
and scalar in three dimensions, we do not need to consider
scalar combinations associated with 
$R_{\mu \nu \lambda \sigma}$.

In Ho\v{r}ava-Lifshitz gravity there are other scalar quantities 
that generate spatial derivatives up to 6-th order:
\be
{\cal Z}_1 \equiv \nabla_i {\cal R} \nabla^i {\cal R}\,,\qquad
{\cal Z}_2 \equiv \nabla_i {\cal R}_{jk} \nabla^i {\cal R}^{jk}\,. 
\label{Zdef}
\ee
We can also take into account the terms like 
${\cal R}^{j}_{i} {\cal R}^{k}_{j}  {\cal R}^{i}_{k}$ 
and ${\cal R} {\cal R}^{j}_{i} {\cal R}^{i}_{j}$, but they are 
irrelevant to the dynamics of linear scalar perturbations 
on the flat FLRW background.
Hence we do not incorporate those terms in the 
following analysis.

In the original Ho\v{r}ava-Lifshitz gravity \cite{Horava} the space-time 
foliation is preserved by the space-independent reparametrization 
$t \to t'(t)$, so the lapse $N$ is assumed to be a function of time $t$ 
alone (which is called the projectability condition). 
This can be extended to a non-projectable version in which 
the lapse depends on both the spatial coordinate $x^i$ ($i=1,2,3$) 
and $t$ \cite{Sergey}. Since the acceleration $a_i=\nabla_i \ln N$ 
does not vanish in this case, we can also consider
the following scalar combinations:
\ba
& &
\alpha_1 \equiv a_i a^i\,,\qquad
\alpha_2 \equiv a_i \Delta a^i\,,\qquad
\alpha_3 \equiv {\cal R} \nabla_i a^i\,,\nonumber \\
& &
\alpha_4 \equiv a_i \Delta^2 a^i\,,\qquad
\alpha_5 \equiv \Delta {\cal R} \nabla_i a^i\,,
\label{aldef}
\ea
where $\Delta \equiv \nabla_i \nabla^i$. 

The action of general modified gravitational theories that depends on 
the above mentioned scalar quantities is given by 
\be
S=\int d^4 x \sqrt{-g}\,L \left( N, K, \s, {\cal R}, {\cal Z}, {\cal U}, 
{\cal Z}_1, {\cal Z}_2, \alpha_1, \cdots, \alpha_5; t \right),
\label{action}
\ee
where $g$ is a determinant of the metric $g_{\mu \nu}$ and 
$L$ is a Lagrangian. The action (\ref{action}) encompasses 
Horndeski/GLPV theories and Ho\v{r}ava-Lifshitz gravity.
In the following we will present explicit forms of the 
Lagrangians in these theories.

First of all, Horndeski theories are described by the Lagrangian
\ba
L &=& G_2(\phi,X)+G_{3}(\phi,X)\square\phi \nonumber \\
&& +G_{4}(\phi,X)\, R-2G_{4,X}(\phi,X)\left[ (\square \phi)^{2}
-\phi^{;\mu \nu }\phi _{;\mu \nu} \right] \nonumber \\
& &
+G_{5}(\phi,X)G_{\mu \nu }\phi ^{;\mu \nu}
+\frac{1}{3}G_{5,X}(\phi,X)
[ (\square \phi )^{3} \nonumber \\
& &-3(\square \phi )\,\phi _{;\mu \nu }\phi ^{;\mu
\nu }+2\phi _{;\mu \nu }\phi ^{;\mu \sigma }{\phi ^{;\nu}}_{;\sigma}]\,,
\label{Lho}
\ea
where $\square \phi \equiv (g^{\mu \nu} \phi_{;\nu})_{;\mu}$, and
$G_{j}$ ($j=2,\cdots,5$) are functions in terms of a scalar
field $\phi$ and its kinetic energy 
$X=g^{\mu \nu}\partial_{\mu} \phi \partial_{\nu} \phi$, 
$R$ and $G_{\mu\nu}$ are the Ricci scalar and 
the Einstein tensor in four dimensions, respectively.
Here and in the following, a lower index of $L$ 
denotes the partial derivatives with respect to the scalar quantities 
represented in the index, e.g., $G_{j,X} \equiv\partial G_{j}/\partial X$.
In unitary gauge we have $\phi=\phi(t)$ and $X=-\dot{\phi}(t)^2/N^2$, 
where a dot represents a derivative with respect to $t$. 
Hence the dependence of $\phi$ and $X$ in the action (\ref{Lho})
is interpreted as that of the lapse $N$ and the time $t$. 
In fact, we can express the Lagrangian (\ref{Lho}) 
of the form \cite{Piazza,Gergely,Gleyzes}
\ba
L &=&
A_2(N,t)+A_3(N,t)K+A_4(N,t) (K^2-{\cal S}) \nonumber \\
& & +B_4(N,t){\cal R} +A_5(N,t) K_3 \nonumber \\
& & +B_5(N,t) \left( {\cal U}-K {\cal R}/2 \right)\,,
\label{LH}
\ea
where $K_3=K^3-3KK_{\mu \nu}K^{\mu \nu}
+2K_{\mu \nu}K^{\mu \lambda}{K^{\nu}}_{\lambda}$. 
Horndeski theories have the following correspondence
\ba
& & A_2=G_2-XF_{3,\phi}\,,\label{A2}\\
& & A_3=2(-X)^{3/2}F_{3,X}-2\sqrt{-X}G_{4,\phi}\,,\label{A3}\\
& & A_4=-G_4+2XG_{4,X}+XG_{5,\phi}/2\,,\label{A4} \\
& & B_4=G_4+X(G_{5,\phi}-F_{5,\phi})/2\,,\label{B4} \\
& & A_5=-(-X)^{3/2}G_{5,X}/3\,,\\
& & B_5=-\sqrt{-X}F_{5}\,,\label{B5}
\label{AB}
\ea
where $F_3$ and $F_5$ are auxiliary functions satisfying
$G_3=F_3+2XF_{3,X}$ and $G_{5,X}=F_5/(2X)+F_{5,X}$.
{}From Eqs.~(\ref{A4})-(\ref{B5}) the following 
two relations hold
\be
A_4=2XB_{4,X}-B_4\,,\qquad
A_5=-XB_{5,X}/3\,,
\label{ABcon}
\ee
under which the number of 6 independent functions reduces to 4.

GLPV \cite{Gleyzes} generalized Horndeski theories in such a way 
that the coefficients $A_4$, $B_4$, $A_5$, and $B_5$ are not 
necessarily related to each other. 
The general action (\ref{action}) can incorporate 
both Horndeski and GLPV theories described by the 
Lagrangian (\ref{LH}).

The action (\ref{action}) also covers Ho\v{r}ava-Lifshitz gravity 
given by the Lagrangian
\ba
\hspace{-0.5cm}
L &=&
\frac{M_{\rm pl}^2}{2} ( {\cal S}-\lambda K^2
+{\cal R}+\eta_1 \alpha_1) \nonumber \\
\hspace{-0.5cm}
& & 
-\frac12 \left( g_2 {\cal R}^2+g_3 {\cal Z}+
\eta_2 \alpha_2+\eta_3 \alpha_3\right) \nonumber \\
\hspace{-0.5cm}
& & 
-\frac{1}{2M_{\rm pl}^2} \left( g_4 {\cal Z}_1+g_5 {\cal Z}_2+
\eta_4 \alpha_4+\eta_5 \alpha_5 \right)\,,
\label{lagHo}
\ea
where $M_{\rm pl}=2.435 \times 10^{18}$~GeV is the reduced 
Planck mass, and $\lambda$, $\eta_1,\cdots,\eta_5$, 
$g_2,\cdots,g_5$ are constants.
The original Ho\v{r}ava-Lifshitz gravity \cite{Horava} corresponds 
to the case $\eta_1=\cdots=\eta_5=0$, 
whereas its healthy extension \cite{Sergey} 
involves the dependence of acceleration.

\section{The second-order action for tensor perturbations}
\label{persec} 

%
\subsection{Cosmological perturbations}

The perturbed line element involving the 
four scalar perturbations $\delta N$, 
$\psi$, $\zeta$, $E$, and tensor perturbations $\gamma_{ij}$ can 
be written of the form
\ba
ds^{2} &=&
-(1+2\delta N) dt^{2}+2\partial_i \psi dx^{i}dt \nonumber \\
& &+a^{2}(t)[ (1+2\zeta)\hat{h}_{ij}+2\partial_{i} \partial_{j}E] 
dx^{i}dx^{j}\,,
\label{permet}
\ea
where $a(t)$ is the scale factor, and 
\begin{equation}
\hat{h}_{ij}=\delta_{ij}+\gamma _{ij}
+\frac{1}{2} \delta^{mk} \gamma_{im} \gamma_{kj}\,,
\label{gra}
\end{equation}
with ${\rm det}\,\hat{h}=1$.  The tensor perturbation $\gamma_{ij}$ is
traceless and divergence-free, i.e., $\gamma _{ii}=\partial _{i}\gamma
_{ij}=0$.  The last term on the r.h.s.\ 
of Eq.~(\ref{gra}) was introduced for the simplification of
calculations, but it does not affect the second-order action of
tensor modes \cite{Maldacena}.

Under the infinitesimal coordinate transformation $t \to t+\delta t$ and 
$x^i \to x^i+\delta^{ij}\partial_{j}\delta x$, the metric perturbation $E$ 
transforms as $E \to E-\delta x$.
Throughout the paper we choose the gauge $E=0$ 
to fix the spatial threading $\delta x$.

The field perturbation $\delta \phi$ transforms as 
$\delta \phi \to \delta \phi-\dot{\phi}\,\delta t$ under the gauge 
transformation. In Horndeski and GLPV theories the 
unitary gauge $\delta \phi=0$ is chosen to fix the time slicing $\delta t$.
In the projectable version of
Ho\v{r}ava-Lifshitz gravity \cite{Horava}, the lapse $N$ is a function
of $t$ alone and hence $\delta N=0$.  In the non-projectable
Ho\v{r}ava-Lifshitz gravity \cite{Sergey}, there is no such
restriction for the gauge choice.
The different gauge choices associated with the temporal 
coordinate transformation do not affect the second-order action of 
tensor perturbations presented later.

On the flat FLRW background described by the line element
$ds^2=-dt^2+a^2(t) \delta_{ij}dx^idx^j$, the extrinsic curvature and
the intrinsic curvature are given, respectively, by
$\bar{K}_{ij}=H\bar{h}_{ij}$ and $\bar{{\cal R}}_{ij}=0$, 
where a bar represents the background values and 
$H=\dot{a}/a$ is the Hubble parameter.
Then, the scalar quantities appearing in the
Lagrangian $L$ of Eq.~(\ref{action}) are $\bar{N}=1$, $\bar{K}=3H$,
$\bar{\s}=3H^{2}$, $\bar{{\cal R}}=\bar{\Z}=\bar{\U}=0$,
$\bar{\Z}_1=\bar{\Z}_2=0$, and
$\bar{\alpha}_1=\bar{\alpha}_2=\cdots=\bar{\alpha}_5=0$.

Expanding the action (\ref{action}) up to second order in scalar
perturbations for the spatial gauge choice $E=0$, we can obtain the
equations of motion for the background and linear scalar perturbations
without fixing the temporal gauge.  Varying the first-order perturbed
action with respect to $\delta N$ and $\delta \sqrt{h}$, respectively,
the background equations are given by \cite{Piazza,Gergely,Kase14}
\begin{eqnarray}
&&\bar{L}+L_{,N}-3H{\cal F}=0\,,
\label{back1} \\
&&\bar{L}-\dot{{\cal F}}-3H{\cal F}=0\,,
\label{back2} 
\end{eqnarray}
where 
\be
{\cal F} \equiv L_{,K}+2HL_{,\s}\,.
\ee
The linear scalar perturbation equations derived by varying the
second-order action in terms of $\delta N$, $\psi$, and $\zeta$ are
presented in Ref.~\cite{Kase14}.

\subsection{The second-order tensor action}

Let us derive the second-order action of Eq.~(\ref{action}) for
tensor perturbations. 
Regarding the extrinsic curvature, tensor modes 
satisfy the relations $K=3H$ and $\delta
K^{i}_{j}=\delta^{ik}\dot{\gamma}_{kj}/2$.  
Up to first order, the
three-dimensional Ricci tensor reads
\be
{\cal R}_{ij}=-\frac12 a^2 \Delta \gamma_{ij}\,.
\ee
The three-dimensional Ricci scalar from tensor perturbations is 
a second-order quantity, which is given by 
\be
{\cal R}=\frac{1}{4} \delta^{ik} \delta^{jl} \gamma_{ij} 
\Delta \gamma_{kl}\,.
\ee
Then the quantity ${\cal Z}_1$ is fourth-order in perturbations.

On using the above relations, the second-order action
for tensor modes reduces to
$S_h^{(2)}=\int d^4 x\,a^3L_h^{(2)}$, where 
$L_h^{(2)}=L_{,{\cal S}} \delta K^{i}_{j} 
\delta K^{j}_{i}+{\cal E} {\cal R}
+L_{,{\cal Z}} {\cal R}^{i}_{j} {\cal R}^{j}_{i}
+L_{,{\cal Z}_2} {\cal Z}_2$ with 
\be
{\cal E} \equiv L_{,\cal R}+\frac12 \dot{L_{,\cal U}}
+\frac32 H L_{,\cal U}\,.
\ee
More explicitly, it is given by 
\be
S_{h}^{(2)}= 
\int d^4 x\,\frac{a^3}{4} \delta^{ik} \delta^{jl}
\left( L_{,{\cal S}} \dot{\gamma}_{ij} \dot{\gamma}_{kl}
+\gamma_{ij} {\cal O}_t \gamma_{kl} \right)\,,
\label{Sh2}
\ee
where 
\be
{\cal O}_t \equiv {\cal E} \Delta+L_{,{\cal Z}}\Delta^2
-L_{,{{\cal Z}_2}}\Delta^3\,.
\ee
Note that there are no contributions to $S_{h}^{(2)}$ from 
the scalars (\ref{aldef}). 
The condition for avoiding the tensor ghost 
corresponds to $L_{,\cal S}>0$.

Varying the action (\ref{Sh2}) with respect to $\gamma_{ij}$, 
we obtain the equation of motion 
\ba
& &
\ddot{\gamma}_{ij}+\left( 3H+\frac{\dot{L_{,\cal S}}}{L_{,\cal S}} 
\right)\dot{\gamma}_{ij}-c_t^2 \Delta \gamma_{ij} \nonumber \\
& & 
-\frac{L_{,{\cal Z}}}{L_{,\cal S}} \Delta^2 \gamma_{ij}
+\frac{L_{,{\cal Z}_2}}{L_{,\cal S}} \Delta^3 \gamma_{ij}=0\,,
\label{teneq}
\ea
where 
\be
c_t^2 \equiv \frac{{\cal E}}{L_{,\cal S}}\,.
\ee
In the absence of spatial derivatives higher than second order, 
$c_t$ exactly corresponds to the propagation speed of gravitational 
waves. In order to avoid the small-scale Laplacian 
instability in this case, we require that $c_t^2>0$. 

General Relativity corresponds to $G_4=M_{\rm pl}^2/2$ 
and $G_5=0$ in the Horndeski Lagrangian (\ref{Lho}), i.e., 
$-A_4=B_4=M_{\rm pl}^2/2$ and $A_5=B_5=0$ 
in Eq.~(\ref{LH}). 
In this case we have $L_{,{\cal S}}=M_{\rm pl}^2/2$,
${\cal E}=M_{\rm pl}^2/2$, $c_t^2=1$, $L_{,{\cal Z}}=0$, and 
$L_{,{\cal Z}_2} =0$, so
Eq.~(\ref{teneq}) reduces to
$\ddot{\gamma}_{ij}+3H\dot{\gamma}_{ij}
-\Delta \gamma_{ij}=0$.

\section{The inflationary tensor modes}
\label{gwspesec}

In this section we derive the power spectrum of tensor perturbations
generated during inflation. 

\subsection{The power spectrum in Fourier space}

We expand the tensor perturbation 
$\gamma_{ij} ({\bm x}, \tau)$ into the Fourier series as
$\gamma_{ij} ({\bm x}, \tau)=\int \frac{d^3k}{(2\pi)^{3/2}}
e^{i {\bm k} \cdot {\bm x}}\hat{\gamma}_{ij}({\bm k}, \tau)$, where
\be
\hat{\gamma}_{ij}({\bm k}, \tau)= \sum_{\lambda=+,\times} 
\hat{h}_{\lambda} ({\bm k}, \tau) e_{ij}^{(\lambda)} ({\bm k})\,.
\label{gamex1}
\ee
Here, ${\bm k}$ is a comoving wavenumber, 
$\tau \equiv \int a^{-1} dt$ is the conformal time, 
and $e_{ij}^{(\lambda)}$ ($\lambda=+, \times$) are 
symmetric polarization tensors. 
The polarization tensors are transverse ($k_je_{ij}^{(\lambda)}=0$) 
and traceless ($e_{ii}^{(\lambda)}=0$) with the normalization satisfying 
$e^{(\lambda)}_{ij} (\bm{k})  e^{*(\lambda')}_{ij} (\bm{k}) 
= \delta_{\lambda \lambda'}$.
We write the Fourier mode $\hat{h}_{\lambda} ({\bm k}, \tau)$
of the form
\be
\hat{h}_{\lambda} ({\bm k}, \tau)=
h_{\lambda} (k, \tau) a_{\lambda} ({\bm k})+
h_{\lambda}^* (k, \tau) a_{\lambda}^{\dagger} (-{\bm k})\,,
\label{gamex2}
\ee
where the annihilation and creation operators
$a_{\lambda} ({\bm k})$ and $a_{\lambda}^{\dagger} 
({\bm k}')$ obey the commutation relation 
$[a_{\lambda} ({\bm k}),a_{\lambda'}^{\dagger} 
({\bm k}')]=\delta_{\lambda \lambda'}\delta^{(3)}
({\bm k}-{\bm k}')$. 

On the quasi de-Sitter background the conformal time 
is given by $\tau \simeq -1/(aH)$, so that the asymptotic 
past and future correspond to $\tau \to -\infty$ and 
$\tau \to 0$, respectively. The tensor power spectrum 
${\cal P}_h (k)$ is defined by the vacuum expectation value 
of $\hat{\gamma}_{ij}$ in the $\tau \to 0$ limit, as 
$\langle 0| \hat{\gamma}_{ij} ({\bm k}_1, 0) 
 \hat{\gamma}_{ij} ({\bm k}_2, 0)  |0
\rangle=(2\pi^2/k_1^3)\delta^{(3)} 
({\bm k}_1+{\bm k}_2) {\cal P}_h (k_1)$.
On using Eqs.~(\ref{gamex1}) and (\ref{gamex2}), 
it follows that 
\be
{\cal P}_h (k)=\frac{k^3}{2\pi^2} 
\left( |h_+(k,0)|^2+|h_\times (k,0)|^2 \right)\,.
\label{powerspe}
\ee
\subsection{Equation of motion for a canonical field}

A canonically normalized field $v_{\lambda} (k, \tau)$ is defined by 
\be
v_{\lambda} (k, \tau)\equiv z h_{\lambda}(k, \tau)\,,\qquad
z \equiv a \sqrt{L_{,\cal S}/2}\,.
\label{vz}
\ee
Then the kinetic term in the action (\ref{Sh2}) can be 
expressed as $S_K=\int d\tau d^3x 
\sum_{\lambda=+,\times}|v_{\lambda}'|^2/2$, where 
a prime represents a derivative with respect to $\tau$. 
{}From Eq.~(\ref{teneq}) each Fourier component 
$v_{\lambda} (k,\tau)$ obeys the equation of motion
\be
v_{\lambda}''+\left[ \mathcal{K}(k,\tau) -\frac{z''}{z} 
\right]v_{\lambda}=0\,,
\label{veq}
\ee
where the function $\mathcal{K}(k,\tau)$ is defined as
\be
\mathcal{K}(k,\tau) \equiv 
c_t^2 k^2 \left( 1+c_1 \frac{k^2}{a^2M_{\rm pl}^2}
+c_2 \frac{k^4}{a^4M_{\rm pl}^4} \right)\,, 
\label{keff}
\ee
and 
\be
c_1 \equiv -\frac{L_{,\cal Z}M_{\rm pl}^2}{{\cal E}}\,,\qquad
c_2 \equiv -\frac{L_{,{\cal Z}_2}M_{\rm pl}^4}{{\cal E}}\,.
\ee

In the context of low-energy effective field theories, we will discuss
the case where $\mathcal{K}(k,\tau)\simeq c_t^2k^2$, 
such that the linear form of the dispersion relation, $\omega=c_t
k$, is not modified by the nonlinear terms in Eq.~(\ref{keff}) 
well below the cut-off of the theories.
Otherwise, we would need to know the
UV-completion of the theories, or our treatment would break down. In
fact, the non-linear terms are suppressed for the physical wavenumber
$k_{\mathrm{phys}}=k/a$ much below the cut-off value
$k_{\mathrm{phys}}^{\mathrm{max}}\equiv M_{\rm pl}/|c_1|^{1/2}$ or
$M_{\rm pl}/|c_2|^{1/4}$.  In Ho\v{r}ava-Lifshitz gravity
\cite{Horava} and in the trans-Planckian physics studied in
Refs.~\cite{Martin,Niemeyer,Kowalski,Tanaka,Star},
$k_{\mathrm{phys}}^{\mathrm{max}}$ is close to $M_{\rm pl}$, i.e.,
$|c_1| \sim |c_2| \sim O(1)$.  In the EFT approach to inflation
advocated by Weinberg \cite{Weinberg}, the cut-off scale is
slightly smaller than $M_{\rm pl}$, say $\sqrt{\epsilon}M_{\rm pl}$,
where $\epsilon=-\dot{H}/H^2$ is the slow-roll parameter typically of
the order of $0.01$.

Since we have the application to Ho\v{r}ava-Lifshitz gravity and 
the trans-Planckian physics in mind, 
we shall focus on the situation in which the cut-off
scale $k_{\mathrm{phys}}^{\mathrm{max}}$ is much larger than 
the Hubble parameter $H$ during inflation. 
In this case the Hubble radius crossing occurs in the
linear regime of the dispersion relation (i.e., $\mathcal{K} \simeq c_t^2 k^2$),
so that the second and third terms in the parenthesis of
Eq.~(\ref{keff}) are regarded as small corrections to the first term.
In other words, the parameter defined by
\be
\sigma \equiv \frac{c_1 H_k^2}{M_{\rm pl}^2}
\label{sigma}
\ee
is much smaller than 1, where $H_k$ is the Hubble 
parameter at $c_tk=aH$. 
Under this condition the EFT approach to inflation 
can be justified.

According to the previous discussion, we will solve Eq.\ (\ref{veq})
iteratively, and write its solution in the form 
\be
v_{\lambda}=v_{\lambda}^{(0)}+v_{\lambda}^{(1)}\,,
\ee
where the leading-order perturbation $v_{\lambda}^{(0)}$
obeys the equation of motion 
\be
v_{\lambda}^{(0)''}+\left( c_t^2 k^2 -\frac{z''}{z} \right)
v_{\lambda}^{(0)}=0\,.
\label{v0eq}
\ee
The field $v_{\lambda}^{(1)}$ induced by the nonlinear 
corrections to Eq.~(\ref{keff}) satisfies
\ba
& &
v_{\lambda}^{(1)''}+\left( c_t^2 k^2 -\frac{z''}{z} \right)
v_{\lambda}^{(1)} \nonumber \\
& &
=-c_t^2 \frac{k^4}{a^2M_{\rm pl}^2}
\left( c_1 + c_2 \frac{k^2}{a^2M_{\rm pl}^2}\right)
v_{\lambda}^{(0)}.
\label{v1eq}
\ea
In order to solve Eq.~(\ref{v0eq}), we take into account the slow-roll
inflationary corrections to the leading-order solution on the de
Sitter background \cite{Stewart}.  We then substitute the
leading-order solution into Eq.~(\ref{v1eq}) to obtain an iterative
solution of $v_{\lambda}^{(1)}$.

\subsection{Solutions to the tensor equations of motion}

In the following we consider the situation in which 
the parameters defined by 
\be
\epsilon \equiv -\frac{\dot{H}}{H^2}\,,\qquad
\epsilon_{{\cal S}} \equiv \frac{\dot{L_{,\cal S}}}{HL_{,{\cal S}}}\,,
\qquad 
s \equiv \frac{\dot{c_t}}{Hc_t}
\ee
are much smaller than unity during inflation. 
The smallness of $\epsilon$ comes from the quasi de Sitter 
background. Dividing Eq.~(3.4) by $2H^2L_{,\cal S}$,
the term $\epsilon_{\cal S}$ appears in addition to $\epsilon$.
Hence $\epsilon_{\cal S}$ is at most the same order 
as $\epsilon$. 

The tensor propagation speed 
square in Horndeski theories can be estimated as 
$c_t^2=1+O(\epsilon)$, so the parameter $s$ is of the 
order of $\epsilon^2$ (see Sec.~\ref{Hoper}). In GLPV theories, $c_t^2$ 
generally differs from 1. 
As we will see in Sec.~\ref{GLPVper}, it is possible to obtain
the Einstein frame with $c_t^2$ equivalent to 1 
under the so-called disformal transformation.
Provided that the cosmological background in the Einstein frame 
is quasi de Sitter, we will show that the 
variation of $c_t^2$ in the original frame is small, i.e., 
$|s| \ll 1$.

The quantity $z''/z$, up to next to leading-order corrections, 
can be estimated as 
\be
\frac{z''}{z}=2(aH)^2 \left( 1-\frac12 \epsilon
+\frac34 \epsilon_{\cal S} \right)\,.
\ee
Introducing a dimensionless variable
\be
y \equiv \frac{c_tk}{aH}\,,
\ee
its time derivative obeys $y'=-aHy(1-\epsilon-s)$. 
Then, Eq.~(\ref{v0eq}) can be expressed as
\ba
& &
(1-2\epsilon-2s) y^2\,\frac{d^2 v_{\lambda}^{(0)}}{dy^2}
-sy \frac{dv_{\lambda}^{(0)}}{dy} \nonumber \\
& &+\left( y^2-2+\epsilon-\frac32 \epsilon_{\cal S} \right)
v_{\lambda}^{(0)}=0\,.
\label{vlameq}
\ea
Here and in the following, we drop contributions of the slow-roll
corrections of the order of $\epsilon^2$.  In other words, we deal
with the first-order slow-roll parameters as constants.

The solution to Eq.~(\ref{vlameq}), after neglecting
non-linear terms in the slow-roll parameters, is given by
\ba
v_{\lambda}^{(0)} (y)
&=& y^{(1+s)/2} \{ \alpha_k 
H_{\nu}^{(1)}[(1+\epsilon+s)y] \nonumber \\
& &~~~~~~~~~~~
+\beta_k H_{\nu}^{(2)}[(1+\epsilon+s)y]\}\,,
\label{vso} 
\ea
where $\alpha_k$ and $\beta_k$ are integration constants,
$H_{\nu}^{(1)}(x)$ and $H_{\nu}^{(2)}(x)$ are Hankel functions of the
first and second kinds respectively, and
\be
\nu=\frac32+\epsilon+\frac12 \epsilon_{\cal S}
+\frac32 s\,.
\ee
The Bunch-Davies vacuum corresponds to the choice $\beta_k=0$. On
using the property $H_{\nu}^{(1)} (x \gg 1) \simeq -\sqrt{2/(\pi
  x)}\,e^{i[x+(3-2\nu)\pi/4]}$, the solution in the asymptotic past
reads
\be
v_{\lambda}^{(0)}(y \gg 1) \simeq -\alpha_k \sqrt{\frac{2}{\pi}} 
\frac{y^{s/2}}{\sqrt{1+\epsilon+s}}
e^{i [(1+\epsilon+s)y+(3-2\nu)\pi/4 ]}\,.
\ee
 The coefficient $\alpha_k$ is determined by the Wronskian condition
$v_{\lambda}^{(0)}v_{\lambda}^{(0)*'}
-v_{\lambda}^{(0)*}v_{\lambda}^{(0)'}=i$, 
such that (up to second order in the slow-roll parameters)
\be
\alpha_k =-\frac14 \sqrt{\frac{\pi}{c_{tk}k}}\,(2+\epsilon+s)\,,
\label{alde}
\ee
where $c_{tk}$ is the value of $c_t$ at $c_tk=aH$ (i.e., at $y=1$).
For the derivation of Eq.~(\ref{alde}) we used the property that any
time-dependent function $f(\tau)$ on the quasi de Sitter background
can be expanded around $y=1$ (denoted by the subscript $k$), as
$f(\tau)=f(\tau_k)-(\dot{f}/H_k) \ln (\tau/\tau_k)$ \cite{Chen}.  For
$\mu$ much smaller than 1 the quantity $y$ is also expanded as
$y^{\mu} \simeq 1+\mu \ln (\tau/\tau_k)$, so the variation of $c_t$,
$H$, and $L_{,\cal S}$ can be quantified as
\be
c_t=c_{tk} y^{-s}\,,\quad
H=H_k y^{\epsilon}\,,\quad
L_{,\cal S}=L_{,{\cal S}k}y^{-\epsilon_{\cal S}}\,.
\label{ctre}
\ee

Substituting Eq.~(\ref{alde}) and $\beta_k=0$ into Eq.~(\ref{vso}), we
obtain
\ba
v_{\lambda}^{(0)} (y) 
&=& -\frac{\sqrt{\pi}}{2} \frac{aH}{(c_tk)^{3/2}} 
\left( 1+\frac12 \epsilon+\frac12 s \right) \nonumber \\
&& \times y^{3/2} H_{\nu}^{(1)}[(1+\epsilon+s)y]\,. 
\label{vlamso}
\ea
Using the property $H_{\nu}^{(1)}(x \to 0)=-(i/\pi)\Gamma(\nu)
(x/2)^{-\nu}$ and the relations (\ref{ctre}), the solution
$h_{\lambda}^{(0)}(y)=v_{\lambda}^{(0)}(y)/z$ long after the Hubble
radius crossing ($y \to 0$) reduces to
\be
h_{\lambda}^{(0)} (0)=i \frac{H_k}{\sqrt{2\pi L_{,{\cal S}k}}}
\frac{2^\nu \Gamma(\nu)}{(c_{tk}k)^{3/2}}
(1-\epsilon-s)\,.
\label{hlambda}
\ee
Expanding the function $2^{\nu}\Gamma (\nu)$ around $\nu=3/2$, it
follows that
\ba
\hspace{-1.3cm}
& &h_{\lambda}^{(0)} (0)= 
i \frac{H_k}{\sqrt{L_{,{\cal S}k}}}
\frac{1}{(c_{tk}k)^{3/2}} \biggl[ 1+(1-\gamma-\ln 2) \epsilon 
\nonumber \\
\hspace{-1.3cm}
& &~~~~~~~~~~
+\frac12 (2-\gamma-\ln 2)\epsilon_{\cal S}
+\frac12 (4-3\gamma-3\ln 2)s \biggr],
\label{hlambda2}
\ea
where $\gamma=0.5772...$ is the Euler-Mascheroni constant.

The next step is to derive the solution to Eq.~(\ref{v1eq}) by using
the leading-order solution of Eq.~(\ref{vlamso}) on the de Sitter
background (obtained by setting $\epsilon=\epsilon_{\cal S}=s=0$ 
and $a=-1/(H\tau)$ with $H={\rm constant}$), i.e.,
$v_{\lambda}^{(0)}(\tau)=-i(1+ic_tk \tau)e^{-ic_tk \tau} /[\sqrt{2}
\tau (c_tk)^{3/2}]$. The speed of propagation for this mode, for
large $k$'s, coincides, by construction, with $c_t$, such that this choice
is consistent with the assumption that the corrections do not modify
the standard propagation of tensor modes. 
Integrating Eq.~(\ref{v1eq}) after substitution of the leading-order solution of
$v_{\lambda}^{(0)}$, the resulting particular solution is given by
\ba
\hspace{-0.7cm}
& &v_{\lambda}^{(1)} (\tau)= 
\frac{e^{-ic_tk \tau}}{120\sqrt{2} c_t^4 (c_tk)^{3/2} \tau}
\frac{H^2}{M_{\rm pl}^2}  \nonumber \\
\hspace{-0.7cm}
& &~~~~~~~~~~\times
\biggl[ 5 \biggl( 5c_1c_t^2-7c_2 \frac{H^2}{M_{\rm pl}^2} 
\biggr)(3i-3c_tk \tau-2c_t^3k^3 \tau^3 )  \nonumber \\
\hspace{-0.7cm}
& &~~~~~~~~~~~~
-10i \biggl( 2c_1c_t^2-7c_2 \frac{H^2}{M_{\rm pl}^2} 
\biggr) c_t^4 k^4 \tau^4 \nonumber \\
\hspace{-0.7cm}
& &~~~~~~~~~~~~
-6c_2 \frac{H^2}{M_{\rm pl}^2} (7+2ic_tk \tau)
c_t^5k^5\tau^5 \biggr]\,.
\label{vfirst}
\ea
The correction $v_{\lambda}^{(1)} (\tau)$ has an oscillatory
part $e^{-ic_tk \tau}$, which by construction, follows the
oscillations of the dominant contribution, $v_{\lambda}^{(0)}
(\tau)$. Long after the Hubble radius crossing ($\tau \to 0$), the
perturbation $h_{\lambda}^{(1)}(\tau)=v_{\lambda}^{(1)}(\tau)/z$
approaches
\be h_{\lambda}^{(1)}(0)=-i \frac{H}{8\sqrt{L_{,\cal
      S}}\,(c_tk)^{3/2}} \left( 5c_1 \frac{H^2}{c_t^2M_{\rm pl}^2}
  -7c_2 \frac{H^4}{c_t^4M_{\rm pl}^4} \right).
\label{hlam1}
\ee
Since we are not interested in the next-order solution to 
Eq.~(\ref{hlam1}), we can replace $H$, $c_t$, and 
$L_{,\cal S}$ for $H_k$, $c_{tk}$, and $L_{,{\cal S}k}$, 
respectively.

\subsection{The spectrum of inflationary tensor modes}

The tensor power spectrum is known by substituting
$h_{\lambda}(0)=h_{\lambda}^{(0)}(0)+h_{\lambda}^{(1)}(0)$ into
Eq.~(\ref{powerspe}), as
\ba
\hspace{-0.5cm}
& &
{\cal P}_{h}(k)=\frac{H_k^2}{\pi^2 L_{,{\cal S}k}c_{tk}^3} 
\biggl[ 1-2(C+1)\epsilon-C \epsilon_{\cal S}
-(3C+2)s
\nonumber \\
\hspace{-0.5cm}
& &~~~~~~~~~~~~~~~~~~~~~~~~~~
-\frac54 \frac{\sigma}{c_{tk}^2}
+\frac{7c_2}{4c_1^2}  \frac{\sigma^2}{c_{tk}^4}\biggr]\,,
\label{Phf}
\ea
where $C=\gamma-2+\ln 2=-0.729637...$ and $\sigma$ 
is defined by Eq.~(\ref{sigma}).
The leading-order power spectrum is given by 
${\cal P}_h^{\rm lead}(k)=H_k^2/(\pi^2 L_{,{\cal S}k}c_{tk}^3)$.

The last two terms in the square bracket of Eq.~(\ref{Phf}), which 
correspond to the corrections induced by spatial derivatives 
higher than second order, are suppressed by the factor 
$\sigma \approx H_k^2/(k_{\mathrm{phys}}^{\mathrm{max}})^2$.  
Provided that $\sigma/c_{tk}^2 \ll \epsilon$,
these terms are smaller than the slow-roll corrections.

We introduce the tensor spectral index $n_t$, as
\be
n_t \equiv \frac{d \ln {\cal P}_{h}(k)}{d\ln k}
\biggr|_{c_{t}k=aH}\,.
\ee
On using the property $d\ln k/dt|_{c_tk=aH}=H(1-\epsilon-s)$ 
and defining the following slow-roll parameters 
\be
\eta \equiv \frac{\dot{\epsilon}}{H\epsilon}\,,\qquad
\eta_{\cal S} \equiv \frac{\dot{\epsilon}_{\cal S}}
{H \epsilon_{\cal S}}\,,\qquad
\delta_s \equiv \frac{\dot{s}}{Hs}\,,
\ee
it follows that 
\ba
n_t &=&
-2\epsilon-\epsilon_{\cal S}-3s-2\epsilon^2
-5\epsilon s-\epsilon \epsilon_{\cal S}-\epsilon_{\cal S}s
-3s^2 \nonumber \\
&&-2(C+1) \epsilon \eta-C\epsilon_{\cal S}\eta_{\cal S}
-(3C+2)s \delta_s \nonumber \\
& & +\frac{5}{2c_{tk}^2} \sigma (\epsilon+s)
-\frac{7c_2}{4c_1^2}  \frac{1}{c_{tk}^4}
\sigma^2 (\epsilon+s)\,,
\label{ntf}
\ea
which should be evaluated at $c_tk=aH$.
The leading-order spectral index is given by 
$n_t^{\rm lead}=-2\epsilon-\epsilon_{\cal S}-3s$.

\section{Application to concrete theories}
\label{appsec}

We estimate the inflationary tensor power spectrum 
and its spectral index in concrete modified gravitational 
theories by using the general results derived 
in Sec.~\ref{gwspesec}.

\subsection{Theories with higher-order spatial derivatives}

Let us consider the theories described by the Lagrangian
\ba
L &=&
\frac{M_{\rm pl}^2}{2} ( {\cal S}-\lambda K^2
+{\cal R})+A_2(N,t)+A_3(N,t)K \nonumber \\
& & 
+\frac{M_{\rm pl}^2}{2} \eta_1 \alpha_1 
- \frac12 \left( g_2 {\cal R}^2+g_3 {\cal Z}+
\eta_2 \alpha_2+\eta_3 \alpha_3\right) \nonumber \\
& & 
-\frac{1}{2M_{\rm pl}^{2}}\left( g_4 {\cal Z}_1+g_5 {\cal Z}_2+
\eta_4 \alpha_4+\eta_5 \alpha_5 \right)\,.
\label{lagHo2}
\ea
For $A_2=A_3=0$ this corresponds to the 
Lagrangian (\ref{lagHo}) of Ho\v{r}ava-Lifshitz gravity, 
including both the projectable ($\alpha_i=0$) and 
non-projectable ($\alpha_i \neq 0$) versions.

We take into account the terms $A_2(N,t)$ and 
$A_3(N,t)K$ in Eq.~(\ref{lagHo2}) to realize inflation 
by a scalar degree of freedom.
In fact, the Lagrangian $L=(M_{\rm pl}^2/2)R+
G_2(\phi,X)+G_3(\phi,X) \square \phi$ 
of the kinetic braiding theories \cite{Vikman} reduces to 
Eq.~(\ref{lagHo2}) with $\lambda=1$, 
$A_2=G_2-XF_{3,\phi}$, $A_3=2(-X)^{3/2}F_{3,X}$, 
$\eta_1=\cdots=\eta_5=0$ and $g_2=\cdots=g_5=0$ in unitary gauge, 
where we used the fact that the four-dimensional Ricci scalar
is expressed as $R={\cal S}-K^2+{\cal R}$ 
up to a boundary term. The field $\phi$ is responsible 
for the cosmic acceleration as it happens for k-inflation ($G_3=0$) 
and potential-driven slow-roll inflation ($G_3=0$ and $G_2=-X/2-V(\phi)$).

Since $L_{,\cal S}={\cal E}=M_{\rm pl}^2/2$, $c_1=g_3$, and $c_2=g_5$, 
Eqs.~(\ref{Phf}) and (\ref{ntf}) read
\ba
{\cal P}_{h}(k) &=& \frac{2H_k^2}{\pi^2 M_{\rm pl}^2} 
\left[ 1-2(C+1) \epsilon
-\frac{5}{4}\sigma+\frac{7g_5}{4g_3^2}\sigma^2 \right]\,,
\label{Ph1}
\nonumber \\ \\
n_t &=& -2\epsilon-2\epsilon^2
-2(C+1) \epsilon \eta
+\frac{5}{2} \sigma \epsilon 
-\frac{7g_5}{4g_3^2} \sigma^2 \epsilon\,,
\label{nt1}
\nonumber \\
\ea
where $\sigma=g_3H_k^2/M_{\rm pl}^2$. 
If $g_3=g_5=0$, then the last two terms in Eqs.~(\ref{Ph1}) 
and (\ref{nt1}) vanish. 
In this case, the above tensor power spectrum reduces to 
the one in standard slow-roll inflation \cite{Stewart}. 

The contributions from the terms $A_2(N,t)$ and $A_3(N,t) K$ 
do not directly appear in Eqs.~(\ref{Ph1})-(\ref{nt1}), but they affect 
the tensor power spectrum indirectly through the background 
equations of motion (\ref{back1})-(\ref{back2}).

Since the leading-order spectrum is ${\cal P}_h^{\rm lead} (k)
=2H_k^2/(\pi^2 M_{\rm pl}^2)$, the energy scale of inflation 
is directly known from the measurement of primordial 
gravitational waves. More concretely, we have 
$H_k/M_{\rm pl} \simeq \pi \sqrt{r{\cal P}_{s}(k)/2}$, 
where ${\cal P}_s(k) \simeq 2.2 \times 10^{-9}$ is the 
observed scalar power spectrum \cite{Planck} 
and $r={\cal P}_h^{\rm lead}(k)/{\cal P}_s(k)$ 
is the tensor-to-scalar ratio. On using the observational bound 
$r \lesssim 0.2$ \cite{Planck}, we have that 
$H_k/M_{\rm pl} \lesssim 4 \times 10^{-5}$. 
Hence, for $|g_3|, |g_5| \lesssim 1$, the corrections induced by 
spatial derivatives higher than second order are suppressed 
compared to the slow-roll corrections (typically 
of the order of 0.01).

Provided that $H$ decreases during inflation, the tensor spectrum 
is red-tilted ($n_t \simeq -2\epsilon<0$).
{}From the background Eqs.~(\ref{back1}) and (\ref{back2}), 
we obtain $M_{\rm pl}^2 (3\lambda-1) \dot{H}=A_{2,N}+3HA_{3,N}$. 
If $\lambda>1/3$, then the condition $\dot{H}<0$ translates to 
$A_{2,N}+3HA_{3,N}<0$. In unitary gauge the field kinetic 
energy is given by $X=-N^{-2}\dot{\phi}^2$, so the Hubble parameter 
decreases for $A_{2,X}+3HA_{3,X}<0$.

\subsection{Horndeski theories}
\label{Hoper}

In unitary gauge the Lagrangian (\ref{Lho}) of Horndeski theories 
is equivalent to Eq.~(\ref{LH}) with the relations (\ref{A2})-(\ref{B5}). 
On using the fact that the term $K_3$ is given by 
$K_3=3H ( 2H^2-2KH+K^2-{\cal S})$ up to quadratic order 
in the perturbations on the flat FLRW background, we have 
$L_{,\cal S}=G_4 (1+\epsilon_1)$ and 
${\cal E}=G_4 (1+\epsilon_2)$, where
\ba
\hspace{-0.5cm}
\epsilon_1 &\equiv& -\frac{2XG_{4,X}}{G_4}
-\frac{X G_{5,\phi}}{2G_4}
+\frac{H(-X)^{3/2} G_{5,X}}{G_4}\,,\\
\hspace{-0.5cm}
\epsilon_2 &\equiv& \frac{X G_{5,\phi}}{2G_4} 
-\frac{X G_{5,X} \ddot{\phi}}{G_4}\,.
\ea
The terms $\epsilon_1$ and $\epsilon_2$, which 
involve $X$, work as the slow-roll corrections to 
the leading-order contribution $G_4$. 
In fact, all these terms appear on the r.h.s. of the 
background equation for $\epsilon$ 
(Eq.~(9) of Ref.~\cite{DeTsu}), so 
they are the same order as $\epsilon$.
The tensor propagation speed square is 
given by $c_t^2 \simeq 1-\epsilon_1+\epsilon_2+O(\epsilon^2)$, 
and hence $s=\epsilon_2 \eta_2/2-\epsilon_1 \eta_1/2+O(\epsilon^3)$, 
where $\eta_j \equiv \dot{\epsilon}_j/(H \epsilon_j)$ with $j=1,2$. 
In the following we set $G_4=(M_{\rm pl}^2/2)F(\phi,X)$, 
where $F(\phi,X)$ is a dimensionless function with 
respect to $\phi$ and $X$. Then the slow-roll parameter 
$\epsilon_{\cal S}$ can be expressed as 
$\epsilon_{\cal S}=\epsilon_{F}+\epsilon_1 \eta_1
+O(\epsilon^3)$, where $\epsilon_{F} \equiv \dot{F}/(HF)$.

The tensor power spectrum and its spectral index, 
up to next to leading-order terms, read
\ba
{\cal P}_h (k) 
&=& \frac{2H_k^2}{\pi^2 M_{\rm pl}^2 F} 
\left[ 1-2(C+1)\epsilon-C\epsilon_{F}
+\frac{\epsilon_1}2-\frac{3\epsilon_2}{2} \right],
\label{Ph2} \nonumber \\ \\
n_t 
&=& -2\epsilon-\epsilon_{F}-2\epsilon^2-\epsilon \epsilon_{F} 
+\frac12 \epsilon_1 \eta_1-\frac32 \epsilon_2 \eta_2
\nonumber \\
& &-2(C+1) \epsilon \eta-C \epsilon_{F} \eta_{F}\,,
\ea
where $\eta_{F} \equiv \dot{\epsilon}_{F}/(H \epsilon_{F})$.

Compared to Eq.~(\ref{Ph1}),  the leading-order power 
spectrum ${\cal P}_h^{\rm lead} (k)=2H_k^2/(\pi^2 M_{\rm pl}^2 F)$ 
of Eq.~(\ref{Ph2}) is divided by the term $F$. 
This term is associated with the conformal factor $\Omega^2$ 
under the transformation 
$\hat{g}_{\mu \nu}=\Omega^2 (\phi, X)g_{\mu \nu}$. 
In the following we study the case in which the conformal factor 
depends on $\phi$ alone, i.e., on $t$ in unitary gauge. 
This assumption is justified provided that the $X$ dependence 
in $\Omega^2$ works only as slow-roll corrections to the 
leading-order $\phi$-dependent term.
Under the conformal transformation 
$\hat{g}_{\mu \nu}=\Omega^2(t)g_{\mu \nu}$, 
the coefficients $A_4$ and $B_4$ in Eq.~(\ref{LH}) 
transform, respectively, as \cite{GlHa}
\ba
\hat{A}_4 &=&\Omega^{-2} A_4 
\left( 1-\frac{3\dot{\Omega}}{N\Omega}\frac{A_5}{A_4} \right)\,,
\label{A4tra} \\
\hat{B}_4 &=&\Omega^{-2} B_4 
\left( 1+\frac{\dot{\Omega}}{2N\Omega}\frac{B_5}{B_4} \right)\,,
\label{B4tra}
\ea
where $A_4=-G_4[1+O(\epsilon)]$ and 
$B_4=G_4[1+O(\epsilon)]$ from Eqs.~(\ref{A4}) and (\ref{B4}). 
Since the second terms in the parentheses of Eqs.~(\ref{A4tra}) 
and (\ref{B4tra}) can be regarded as slow-roll corrections, 
we have $\hat{A}_4=-\Omega^{-2}G_4 [1+O(\epsilon)]$ 
and $\hat{B}_4=\Omega^{-2}G_4 [1+O(\epsilon)]$.
Choosing the conformal factor $\Omega^2=
2G_4/M_{\rm pl}^2=F$, it follows that 
$\hat{A}_4=-(M_{\rm pl}^2/2)[1+O(\epsilon)]$ and 
$\hat{B}_4=(M_{\rm pl}^2/2)[1+O(\epsilon)]$. 

Under the conformal transformation 
$\hat{g}_{\mu \nu}=\Omega^2(t)g_{\mu \nu}$, 
the structure of the Lagrangian (\ref{LH}) is preserved 
with the modified leading-order coefficients 
$\hat{A}_2=\Omega^{-4} A_2$, 
$\hat{A}_3=\Omega^{-3}A_3$, 
$\hat{A}_5=\Omega^{-1}A_5$, and 
$\hat{B}_5=\Omega^{-1}B_5$ in the presence 
of slow-roll corrections (involving the derivative 
$\dot{\Omega}/(N\Omega)$) \cite{GlHa}. 
This means that, for the choice $\Omega^2=F$, 
the leading-order tensor spectrum in the transformed (Einstein)
frame can be derived by setting $L_{,\cal S}=M_{\rm pl}^2/2$ 
and $c_t=1$ in Eq.~(\ref{Phf}), i.e., $\hat{{\cal P}}_h^{\rm lead} (k)=
2\hat{H}_k^2/(\pi^2 M_{\rm pl}^2)$.
Since the Hubble parameters in two frames are related to 
each other as $\hat{H}=[H+\dot{F}/(2NF)]/\sqrt{F}$, 
the spectrum $\hat{{\cal P}}_h^{\rm lead} (k)$ 
is equivalent to 
${\cal P}_h^{\rm lead} (k)=2H_k^2/(\pi^2 M_{\rm pl}^2 F)$
at leading order in slow-roll.
Provided that the null energy condition is not violated in the 
Einstein frame the Hubble parameter $\hat{H}$ decreases, 
in which case the tensor power spectrum is red-tilted.

The above properties can be notably seen in the Higgs inflationary scenario 
with the scalar-field potential $V(\phi)=(\lambda/4)(\phi^2-v^2)^2$
and the function $F=1+\zeta \phi^2/M_{\rm pl}^2$, 
where $\zeta$ is a non-minimal coupling \cite{Higgs} 
(see also Refs.~\cite{Unruh}).
In order to realize the self-coupling $\lambda$ of the order 
of 0.1, the non-minimal coupling is constrained to be 
$\zeta=O(10^4)$ from the CMB normalization.
For $\zeta \gg 1$ the quantity $F$ is related to the number of 
e-foldings $N_e$ from the end of inflation, 
as $F \simeq 4N_e/3$ \cite{Komatsu}, which 
is much larger than 1 on scales relevant to the CMB anisotropies. 
The action in the Einstein frame is characterized by a canonically 
normalized field with the potential $\hat{V}=V(\phi)/F^2$ \cite{Maeda}, 
in which case the tensor spectrum 
${\cal P}_h^{\rm lead} (k)=
2\hat{H}_k^2/(\pi^2 M_{\rm pl}^2)$ is red-tilted due to 
the decrease of $\hat{H}$.

\subsection{GLPV theories}
\label{GLPVper}

Let us proceed to the GLPV theories in unitary gauge, i.e., 
the Lagrangian (\ref{LH}). In this case
the functions $L_{,\cal S}$ and ${\cal E}$ 
are given by $L_{,\cal S}=-A_4 (1+\epsilon_1)$ and 
${\cal E}=B_4(1+\epsilon_2)$ respectively, where
\be
\epsilon_1=\frac{3HA_5}{A_4}\,,\qquad
\epsilon_2=\frac{\dot{B}_5}{2B_4}\,.
\ee
Provided that $\epsilon_1$ and $\epsilon_2$ are 
regarded as slow-roll corrections to the leading-order 
terms of $L_{,\cal S}$ and ${\cal E}$, we have
$c_t^2=-(B_4/A_4)(1-\epsilon_1+\epsilon_2$). 
The difference from Horndeski theories is that 
$A_4$ and $B_4$ are not related with each other, 
so $c_t^2$ generally differs from 1.
Then the leading-order tensor spectrum is given by 
\be
{\cal P}_h^{\rm lead}(k)=\frac{H_k^2}
{\pi^2 |A_4|\,c_{tk,{\rm lead}}^3} \,,
\label{tenGLPV}
\ee
where $c_{tk,{\rm lead}}^2=-B_4/A_4$.

We perform the disformal transformation given by 
$\tilde{g}_{\mu \nu}=g_{\mu \nu}+\Gamma(\phi,X) 
\partial_{\mu}\phi \partial_{\nu} \phi$, where 
$\Gamma(\phi,X)$ is a function in terms of 
$\phi$ and $X$ \cite{Beken,Garcia}. 
In Ref.~\cite{GlHa} it was shown that the structure of the 
GLPV action is preserved under this 
transformation\footnote{In the presence of an additional 
matter there is a mixing between the sound 
speeds of the scalar field $\phi$ and matter in GLPV theories 
even for the metric frame minimally coupled to 
matter \cite{Gergely,Gleyzes}. 
The disformal transformation gives rise to a kinetic-type coupling 
of the scalar field with matter in the transformed
frame \cite{Garcia,GlHa}, which 
helps us to understand the origin of such a non-trivial mixing. 
Here we do not take into account an additional matter, as 
we are interested in the application to single-field inflation.}.
The coefficients $A_4$ and $B_4$ in the Lagrangian
(\ref{LH}) are transformed as
\be
\tilde{A}_4=\sqrt{1+\Gamma X}\,A_4\,,\qquad
\tilde{B}_4=\frac{B_4}{\sqrt{1+\Gamma X}}\,.
\ee
In the new frame the tensor propagation speed square is 
given by $\tilde{c}_{t,{\rm lead}}^2=-\tilde{B}_4/\tilde{A}_4
=c_{t,{\rm lead}}^2/(1+\Gamma X)$. 
If we choose the function
\be
\Gamma=-\frac{1-c_{t,{\rm lead}}^2}{X}\,,
\label{Gamcho}
\ee
then it follows that $\tilde{c}_{t,{\rm lead}}^2=1$. 
In this case, the coefficients in Eq.~(\ref{LH}) are 
transformed as $\tilde{A}_2=A_2/c_{t,{\rm lead}}$, 
$\tilde{A}_3=A_3$, $\tilde{A}_4=c_{t,{\rm lead}}A_4$, 
$\tilde{B}_4=B_4/c_{t,{\rm lead}}$, 
$\tilde{A}_5=c_{t,{\rm lead}}^2A_5$, 
and $\tilde{B}_5=B_5$. 
Since $\tilde{c}_{t,{\rm lead}}^2=1$ in the new frame, 
the leading-order spectrum becomes 
${\cal P}^{\rm lead}_h(k)=\tilde{H}_k^2/(\pi^2 |\tilde{A}_4|)$.
If we make the conformal transformation $\hat{g}_{\mu \nu}=
\Omega^2(t) \tilde{g}_{\mu \nu}$ further with 
$\Omega^2=2|\tilde{A}_4|/M_{\rm pl}^2$, 
the resulting leading-order spectrum 
reduces to ${\cal P}_h^{\rm lead} (k)=
2\hat{H}_k^2/(\pi^2 M_{\rm pl}^2)$.

Under the disformal transformation 
$\tilde{g}_{\mu \nu}=g_{\mu \nu}+\Gamma(\phi,X) 
\partial_{\mu}\phi \partial_{\nu} \phi$, the lapse 
function $N$ is generally transformed to 
$\tilde{N}=N\sqrt{1+\Gamma X}$ \cite{GlHa,Tsu14}. 
Setting $N=1$ for the background, the choice of 
$\Gamma$ in Eq.~(\ref{Gamcho}) can be interpreted as 
$\tilde{N}=c_{t,{\rm lead}}$.
The Hubble parameters in the Einstein and original frames 
are related with each other as $\tilde{H}=H/\tilde{N}=H/c_{t,{\rm lead}}$. 
This leads to the relation $\tilde{\epsilon}=\epsilon+s$, 
where $\tilde{\epsilon}=-\dot{\tilde{H}}/(\tilde{N}\tilde{H}^2)$ 
and $s=\dot{c}_{t,{\rm lead}}/(Hc_{t,{\rm lead}})$. 
Provided that the cosmological background in the Einstein frame is 
quasi de Sitter, we have that $\tilde{\epsilon} \ll 1$ 
and hence $|s| \ll 1$. Thus the assumption $|s| \ll 1$ used 
to derive the tensor power spectrum (\ref{Phf}) is justified.

The above discussion shows that the combination of the 
disformal and conformal transformations, 
$\hat{g}_{\mu \nu}=\Omega^2(\phi)g_{\mu \nu}
+\Gamma(\phi,X)\partial_{\mu}\phi \partial_{\nu} \phi$, 
can lead to a metric frame in which the leading-order 
tensor power spectrum is of the standard form that depends 
on the Hubble parameter $\hat{H}_k$ alone.
This conclusion is consistent with the recent results of 
Ref.~\cite{Cre} in which the authors took the EFT approach 
without having the direct connection to particular modified 
gravitational theories.

\section{Conclusions}
\label{conclude} 

We have studied tensor perturbations on the flat FLRW background
for the general action (\ref{action}) that encompasses most of the 
modified gravitational theories proposed 
in the literature--including Horndeski theories, GLPV theories, and 
Ho\v{r}ava-Lifshitz gravity. 
The equation of motion (\ref{teneq}), which follows from 
the second-order action (\ref{Sh2}), involves the spatial derivatives 
higher than second order for the theories where the Lagrangian 
$L$ depends on ${\cal Z}$ or ${\cal Z}_2$.

We derived the inflationary power spectrum of tensor modes
under the condition that the cut-off scale $k_{\mathrm{phys}}^{\mathrm{max}}$ associated 
with the non-linear terms of Eq.~(\ref{keff}) is much larger than 
the Hubble parameter $H_k$ at $c_tk=aH$ during inflation. 
On using the small parameter $\sigma$ of the order of 
$H_k^2/(k_{\mathrm{phys}}^{\mathrm{max}})^2$,
the solution to Eq.~(\ref{veq}) is obtained iteratively on the de Sitter background. 
Taking into account the slow-roll corrections to the 
leading-order solution as well, the resulting tensor power spectrum 
is given by Eq.~(\ref{Phf}) with the spectral index (\ref{ntf}).

The corrections from the higher-order spatial derivatives 
to the leading-order power spectrum are suppressed by 
the factor $\sigma/c_{tk}^2$. This conclusion is consistent 
with the effect of modified trans-Planckian dispersion relations 
on the inflationary power spectrum \cite{Niemeyer,Kowalski,Tanaka,Star}.
For $k_{\mathrm{phys}}^{\mathrm{max}}$ close to $M_{\rm pl}$ and 
for $c_{tk}$ not very much smaller than 1, 
the corrections induced by the spatial derivatives 
higher than second order are smaller than 
the slow-roll corrections arising from the deviation from 
the de Sitter background.

We applied our general formula of the inflationary tensor 
power spectrum to a number of concrete modified 
gravitational theories. For the Lagrangian (\ref{lagHo2}), which 
encompasses kinetic braiding models and Ho\v{r}ava-Lifshitz gravity, 
the leading-order spectrum is directly related to $H_k$, 
as ${\cal P}^{\rm lead}_h(k)=2H_k^2/(\pi^2M_{\rm pl}^2)$. 

In Horndeski theories, where the tensor propagation speed is 1 
at leading-order in slow-roll, ${\cal P}^{\rm lead}_h(k)$ 
involves a dimensionless factor $F=2G_4/M_{\rm pl}^2$ 
in the denominator. Under the conformal transformation 
$\hat{g}_{\mu \nu}=Fg_{\mu \nu}$, the spectrum 
in the Einstein frame simply reduces to
${\cal P}^{\rm lead}_h(k)=2\hat{H}_k^2/(\pi^2M_{\rm pl}^2)$.

In GLPV theories the leading-order tensor spectrum (\ref{tenGLPV}) 
involves the terms $A_4$ and $c_{tk,{\rm lead}}^2=-B_4/A_4$.
We showed that, under the disformal transformation 
$\hat{g}_{\mu \nu}=\Omega^2(\phi)g_{\mu \nu}
+\Gamma(\phi,X)\partial_{\mu}\phi \partial_{\nu} \phi$, 
it is possible to find a frame in which 
$\hat{c}_{tk,{\rm lead}}^2=1$ and $\hat{A}_4=-M_{\rm pl}^2/2$ 
up to slow-roll corrections.
Thus the prediction of inflationary tensor modes is robust 
in that there exists the metric frame in which 
the leading-order spectrum is simply proportional to 
$\hat{H}_k^2$ in a vast class of modified gravitational theories.

\section*{Acknowledgements}

This work is supported by the Grant-in-Aid for Scientific 
Research from JSPS (No.~24540286) 
and by the cooperation programs of Tokyo University of Science 
and CSIC. 



\begin{thebibliography}{99}

\bibitem{Sta80}
A.~A.~Starobinsky,
Phys.\ Lett.\ B {\bf 91}, 99 (1980).

\bibitem{Kazanas}
D.~Kazanas,
Astrophys.\ J.\  {\bf 241} L59 (1980);
K.~Sato, Mon.\ Not.\ R.\ Astron.\ Soc. {\bf 195}, 467 (1981);
Phys.\ Lett.\ {\bf 99B}, 66 (1981);
A.~H.~Guth,
Phys.\ Rev.\ D {\bf 23}, 347 (1981).

\bibitem{newinf}
A.~D.~Linde,
Phys.\ Lett.\ B {\bf 108}, 389 (1982);
A.~Albrecht and P.~J.~Steinhardt,
Phys.\ Rev.\ Lett.\  {\bf 48}, 1220 (1982);
A.~D.~Linde,
Phys.\ Lett.\ B {\bf 129}, 177 (1983).

\bibitem{GWper}
A.~A.~Starobinsky,
JETP Lett.\  {\bf 30}, 682 (1979);
V.~A.~Rubakov, M.~V.~Sazhin and A.~V.~Veryaskin,
Phys.\ Lett.\ B {\bf 115}, 189 (1982);
R.~Fabbri and M.~d.~Pollock,
Phys.\ Lett.\ B {\bf 125}, 445 (1983).

\bibitem{oldper}
V.~F.~Mukhanov and G.~V.~Chibisov,
JETP Lett.\  {\bf 33}, 532 (1981);
A.~H.~Guth and S.~Y.~Pi,
Phys.\ Rev.\ Lett.\  {\bf 49} (1982) 1110;
S.~W.~Hawking,
Phys.\ Lett.\ B {\bf 115}, 295 (1982);
A.~A.~Starobinsky,
Phys.\ Lett.\ B {\bf 117} (1982) 175;
J.~M.~Bardeen, P.~J.~Steinhardt and M.~S.~Turner,
Phys.\ Rev.\ D {\bf 28}, 679 (1983).

\bibitem{Planck} 
P.~A.~R.~Ade {\it et al.}  [Planck Collaboration],
Astron.\ Astrophys.\  (2014)
[arXiv:1303.5076 [astro-ph.CO]].

\bibitem{QUIET} 
D.~Araujo {\it et al.}  [QUIET Collaboration],
Astrophys.\ J.\  {\bf 760}, 145 (2012)
[arXiv:1207.5034 [astro-ph.CO]].

\bibitem{BICEP} 
P.~A.~R.~Ade {\it et al.}  [BICEP2 Collaboration],
Phys.\ Rev.\ Lett.\  {\bf 112}, 241101 (2014)
[arXiv:1403.3985 [astro-ph.CO]].

\bibitem{Polar} 
P.~A.~R.~Ade {\it et al.}  [The POLARBEAR Collaboration],
Astrophys.\ J.\  {\bf 794}, 171 (2014)
[arXiv:1403.2369 [astro-ph.CO]].

\bibitem{Lite} 
T.~Matsumura  {\it et al.},
Journal of Low Temperature Physics September 2014, Volume 176,
Issue 5-6, pp 733-740
[arXiv:1311.2847 [astro-ph.IM]].

\bibitem{Horndeski} 
G.~W.~Horndeski, Int.\ J.\ Theor.\ Phys.\ 10,
363-384 (1974).

\bibitem{KYY} 
T.~Kobayashi, M.~Yamaguchi and J.~'i.~Yokoyama,
Prog.\ Theor.\ Phys.\  {\bf 126}, 511 (2011)
[arXiv:1105.5723 [hep-th]].

\bibitem{GaoSteer} 
X.~Gao and D.~A.~Steer, 
JCAP \textbf{1112}, 019 (2011) {[}arXiv:1107.2642 {[}astro-ph.CO{]}{]};
A.~De Felice and S.~Tsujikawa,
JCAP {\bf 1104}, 029 (2011)
[arXiv:1103.1172 [astro-ph.CO]].

\bibitem{kinf} 
C.~Armendariz-Picon, T.~Damour and V.~F.~Mukhanov,
Phys.\ Lett.\ B {\bf 458}, 209 (1999)
[hep-th/9904075].

\bibitem{Higgs} 
F.~L.~Bezrukov and M.~Shaposhnikov,
Phys.\ Lett.\ B {\bf 659}, 703 (2008)
[arXiv:0710.3755 [hep-th]].

\bibitem{Germani} 
C.~Germani and A.~Kehagias,
Phys.\ Rev.\ Lett.\  {\bf 105}, 011302 (2010)
[arXiv:1003.2635 [hep-ph]].

\bibitem{Reza} 
A.~De Felice, S.~Tsujikawa, J.~Elliston and R.~Tavakol,
JCAP {\bf 1108}, 021 (2011)
[arXiv:1105.4685 [astro-ph.CO]].

\bibitem{Kuro} 
S.~Tsujikawa, J.~Ohashi, S.~Kuroyanagi and A.~De Felice,
Phys.\ Rev.\ D {\bf 88}, no. 2, 023529 (2013)
[arXiv:1305.3044 [astro-ph.CO]].

\bibitem{Piazza} 
J.~Gleyzes, D.~Langlois, F.~Piazza, and F.~Vernizzi, 
JCAP \textbf{1308}, 025 (2013) 
[arXiv:1304.4840 [hep-th]].

\bibitem{Gleyzes} 
J.~Gleyzes, D.~Langlois, F.~Piazza and F.~Vernizzi,
arXiv:1404.6495 [hep-th].

\bibitem{Lin} 
C.~Lin, S.~Mukohyama, R.~Namba and R.~Saitou,
JCAP {\bf 1410}, no. 10, 071 (2014)
[arXiv:1408.0670 [hep-th]].

\bibitem{GlHa} 
J.~Gleyzes, D.~Langlois, F.~Piazza and F.~Vernizzi,
JCAP {\bf 1502}, 018 (2015)
[arXiv:1408.1952 [astro-ph.CO]].

\bibitem{GaoHa} 
X.~Gao,
Phys.\ Rev.\ D {\bf 90}, 104033 (2014)
[arXiv:1409.6708 [gr-qc]].

\bibitem{KaGe} 
R.~Kase, L.~A.~Gergely and S.~Tsujikawa,
Phys.\ Rev.\ D {\bf 90}, 124019 (2014)
[arXiv:1406.2402 [hep-th]].

\bibitem{Tsu14} 
S.~Tsujikawa,
JCAP {\bf 1504}, 043 (2015)
[arXiv:1412.6210 [hep-th]].

\bibitem{Beken} 
 J.~D.~Bekenstein,
Phys.\ Rev.\ D {\bf 48}, 3641 (1993).

\bibitem{Garcia} 
D.~Bettoni and S.~Liberati,
Phys.\ Rev.\ D {\bf 88}, no. 8, 084020 (2013)
[arXiv:1306.6724 [gr-qc]];
M.~Zumalacarregui and J.~Garcia-Bellido,
Phys.\ Rev.\ D {\bf 89}, 064046 (2014)
[arXiv:1308.4685 [gr-qc]];
M.~Minamitsuji,
Phys.\ Lett.\ B {\bf 737}, 139 (2014)
[arXiv:1409.1566 [astro-ph.CO]].

\bibitem{Horava} 
P.~Ho\v{r}ava,
Phys.\ Rev.\ D {\bf 79}, 084008 (2009)
[arXiv:0901.3775 [hep-th]].

\bibitem{Cremi} 
P.~Creminelli, M.~A.~Luty, A.~Nicolis and L.~Senatore, 
JHEP \textbf{0612}, 080 (2006) [hep-th/0606090].

\bibitem{Cheung} 
C.~Cheung, P.~Creminelli, A.~L.~Fitzpatrick, J.~Kaplan 
and L.~Senatore, 
JHEP \textbf{0803}, 014 (2008) [arXiv:0709.0293 [hep-th]].

\bibitem{Weinberg} 
S.~Weinberg,
Phys.\ Rev.\ D {\bf 77}, 123541 (2008)
[arXiv:0804.4291 [hep-th]].

\bibitem{Smith} 
K.~M.~Smith, L.~Senatore and M.~Zaldarriaga,
JCAP {\bf 0909}, 006 (2009)
[arXiv:0901.2572 [astro-ph.CO]].

\bibitem{Cremi2} 
P.~Creminelli, G.~D'Amico, M.~Musso, J.~Norena and E.~Trincherini,
JCAP {\bf 1102}, 006 (2011)
[arXiv:1011.3004 [hep-th]].

\bibitem{Quin} 
P.~Creminelli, G.~D'Amico, J.~Norena and F.~Vernizzi, 
JCAP \textbf{0902}, 018 (2009) 
[arXiv:0811.0827 [astro-ph]].

\bibitem{Park} 
M.~Park, K.~M.~Zurek and S.~Watson, 
Phys.\ Rev.\ D \textbf{81}, 124008 (2010) 
[arXiv:1003.1722 [hep-th]].

\bibitem{Flanagan} 
J.~K.~Bloomfield and E.~E.~Flanagan, 
JCAP \textbf{1210}, 039 (2012) [arXiv:1112.0303 [gr-qc]].

\bibitem{Battye} 
R.~A.~Battye and J.~A.~Pearson, 
JCAP \textbf{1207}, 019 (2012) [arXiv:1203.0398 [hep-th]].

\bibitem{Mueller} 
E.~M.~Mueller, R.~Bean and S.~Watson, 
Phys.\ Rev.\ D \textbf{87}, 083504 (2013) 
[arXiv:1209.2706 [astro-ph.CO]].

\bibitem{Gubi} 
G.~Gubitosi, F.~Piazza and F.~Vernizzi, 
JCAP \textbf{1302} (2013) 032 [arXiv:1210.0201 [hep-th]].

\bibitem{Bloom} 
J.~K.~Bloomfield, E.~Flanagan, M.~Park and S.~Watson, 
JCAP \textbf{1308}, 010 (2013) [arXiv:1211.7054 [astro-ph.CO]];
J.~Bloomfield, 
JCAP \textbf{1312}, 044 (2013) [arXiv:1304.6712 [astro-ph.CO]].

\bibitem{Piazza2} 
F.~Piazza and F.~Vernizzi, 
Class.\ Quant.\ Grav.\ \textbf{30}, 214007 (2013) [arXiv:1307.4350];
F.~Piazza, H.~Steigerwald and C.~Marinoni,
JCAP {\bf 1405}, 043 (2014)
[arXiv:1312.6111 [astro-ph.CO]].

\bibitem{Silve} 
N.~Frusciante, M.~Raveri and A.~Silvestri,
JCAP {\bf 1402}, 026 (2014)
[arXiv:1310.6026 [astro-ph.CO]];
B.~Hu, M.~Raveri, N.~Frusciante and A.~Silvestri,
Phys.\ Rev.\ D {\bf 89}, 103530 (2014)
[arXiv:1312.5742 [astro-ph.CO]].

\bibitem{Tsuji14} 
S.~Tsujikawa,
Lect.\ Notes Phys.\  {\bf 892}, 97 (2015)
[arXiv:1404.2684 [gr-qc]].

\bibitem{Gergely} 
L.~Gergely and S.~Tsujikawa, 
Phys.\ Rev.\ D \textbf{89}, 064059 (2014) 
[arXiv:1402.0553 [hep-th]].

\bibitem{Gao} 
X.~Gao,
Phys.\ Rev.\ D {\bf 90}, 081501 (2014)
[arXiv:1406.0822 [gr-qc]].

\bibitem{RT14} 
R.~Kase and S.~Tsujikawa,
Phys.\ Rev.\ D {\bf 90}, 044073 (2014)
[arXiv:1407.0794 [hep-th]].

\bibitem{Kase14} 
R.~Kase and S.~Tsujikawa,
Int.\ J.\ Mod.\ Phys.\ D {\bf 23}, 3008 (2014)
[arXiv:1409.1984 [hep-th]].

\bibitem{ins1} 
C.~Charmousis, G.~Niz, A.~Padilla and P.~M.~Saffin,
JHEP {\bf 0908}, 070 (2009)
[arXiv:0905.2579 [hep-th]].

\bibitem{ins2} 
K.~Koyama and F.~Arroja,
JHEP {\bf 1003}, 061 (2010)
[arXiv:0910.1998 [hep-th]].

\bibitem{Sergey}
D.~Blas, O.~Pujolas and S.~Sibiryakov,
Phys.\ Rev.\ Lett.\  {\bf 104}, 181302 (2010)
[arXiv:0909.3525 [hep-th]]. 

\bibitem{Soti} 
A.~Papazoglou and T.~P.~Sotiriou,
Phys.\ Lett.\ B {\bf 685}, 197 (2010)
[arXiv:0911.1299 [hep-th]]

\bibitem{Sergey2}
D.~Blas, O.~Pujolas and S.~Sibiryakov,
Phys.\ Lett.\ B {\bf 688}, 350 (2010)
[arXiv:0912.0550 [hep-th]].

\bibitem{Cre} 
P.~Creminelli, J.~Gleyzes, J.~Norena and F.~Vernizzi,
Phys.\ Rev.\ Lett.\  {\bf 113}, 231301 (2014)
[arXiv:1407.8439 [astro-ph.CO]].

\bibitem{Wands} 
D.~Cannone, G.~Tasinato and D.~Wands,
arXiv:1409.6568 [astro-ph.CO].

\bibitem{ADM} 
R.~L.~Arnowitt, S.~Deser and C.~W.~Misner, 
Phys.\ Rev.\ \textbf{117}, 1595 (1960).

\bibitem{Maldacena} 
J.~M.~Maldacena,
JHEP {\bf 0305}, 013 (2003)
[astro-ph/0210603].

\bibitem{Martin} 
J.~Martin and R.~H.~Brandenberger,
Phys.\ Rev.\ D {\bf 63}, 123501 (2001)
[hep-th/0005209];
R.~H.~Brandenberger and J.~Martin,
Mod.\ Phys.\ Lett.\ A {\bf 16}, 999 (2001)
[astro-ph/0005432].

\bibitem{Niemeyer}
J.~C.~Niemeyer,
Phys.\ Rev.\ D {\bf 63}, 123502 (2001)
[astro-ph/0005533];
J.~C.~Niemeyer and R.~Parentani,
Phys.\ Rev.\ D {\bf 64}, 101301 (2001)
[astro-ph/0101451].

\bibitem{Kowalski}
J.~Kowalski-Glikman,
Phys.\ Lett.\ B {\bf 499}, 1 (2001)
[astro-ph/0006250].

\bibitem{Tanaka}
T.~Tanaka,
astro-ph/0012431.

\bibitem{Star}
A.~A.~Starobinsky,
Pisma Zh.\ Eksp.\ Teor.\ Fiz.\  {\bf 73}, 415 (2001)
[JETP Lett.\  {\bf 73}, 371 (2001)]
[astro-ph/0104043].

\bibitem{Stewart} 
E.~D.~Stewart and D.~H.~Lyth,
Phys.\ Lett.\ B {\bf 302}, 171 (1993)
[gr-qc/9302019].

\bibitem{Chen} 
X.~Chen, M.~x.~Huang, S.~Kachru and G.~Shiu,
JCAP {\bf 0701}, 002 (2007)
[hep-th/0605045];
A.~De Felice and S.~Tsujikawa,
JCAP {\bf 1303}, 030 (2013)
[arXiv:1301.5721 [hep-th]].

\bibitem{Vikman} 
C.~Deffayet, O.~Pujolas, I.~Sawicki and A.~Vikman,
JCAP {\bf 1010}, 026 (2010)
[arXiv:1008.0048 [hep-th]].

\bibitem{DeTsu} 
A.~De Felice and S.~Tsujikawa,
Phys.\ Rev.\ D {\bf 84}, 083504 (2011)
[arXiv:1107.3917 [gr-qc]].

\bibitem{Unruh} 
T.~Futamase and K.~i.~Maeda,
Phys.\ Rev.\ D {\bf 39}, 399 (1989);
R.~Fakir and W.~G.~Unruh,
Phys.\ Rev.\ D {\bf 41}, 1783 (1990).

\bibitem{Komatsu} 
E.~Komatsu and T.~Futamase,
Phys.\ Rev.\ D {\bf 59}, 064029 (1999)
[astro-ph/9901127];
S.~Tsujikawa and B.~Gumjudpai,
Phys.\ Rev.\ D {\bf 69}, 123523 (2004)
[astro-ph/0402185]. 

\bibitem{Maeda} 
K.~i.~Maeda,
Phys.\ Rev.\ D {\bf 39}, 3159 (1989).

\end{thebibliography}
\end{document}